# Building floorspace and stock measurement: A review of global efforts, knowledge gaps, and research priorities☆

Minda Ma [1†, 2], Shufan Zhang [3†], Junhong Liu [1], Ran Yan [3], Weiguang Cai [3], Nan Zhou [2 *], Jinyue Yan [4 *]

1. School of Architecture and Urban Planning, Chongqing University, Chongqing, 400045, PR China
2. Building Technology and Urban Systems Division, Energy Technologies Area, Lawrence Berkeley National Laboratory, Berkeley, CA 94720, United States
3. School of Management Science and Real Estate, Chongqing University, Chongqing, 400045, PR China
4. Department of Building Environment and Energy Engineering, The Hong Kong Polytechnic University, Kowloon, Hong Kong, PR China

- First author: Dr. Minda Ma, Email: maminda@lbl.gov
  Homepage: https://buildings.lbl.gov/people/minda-ma
  http://chongjian.cqu.edu.cn/info/1556/6706.htm

  First author: Ms. Shufan Zhang, Email: zhangshufan@cqu.edu.cn
  †: The two authors contributed equally to the study.

- Corresponding author: Dr. Nan Zhou, Email: nzhou@lbl.gov
  Homepage: https://buildings.lbl.gov/people/nan-zhou

  Corresponding author: Prof. Dr. Jinyue Yan, Email: j-jerry.yan@polyu.edu.hk
  Homepage: https://polyu.edu.hk/beee/people/academic-staff/professor-yan-jinyue-jerry

- Lead contact: Dr. Nan Zhou, Email: nzhou@lbl.gov

Word count of main content: 7890

## GRAPHICAL ABSTRACT

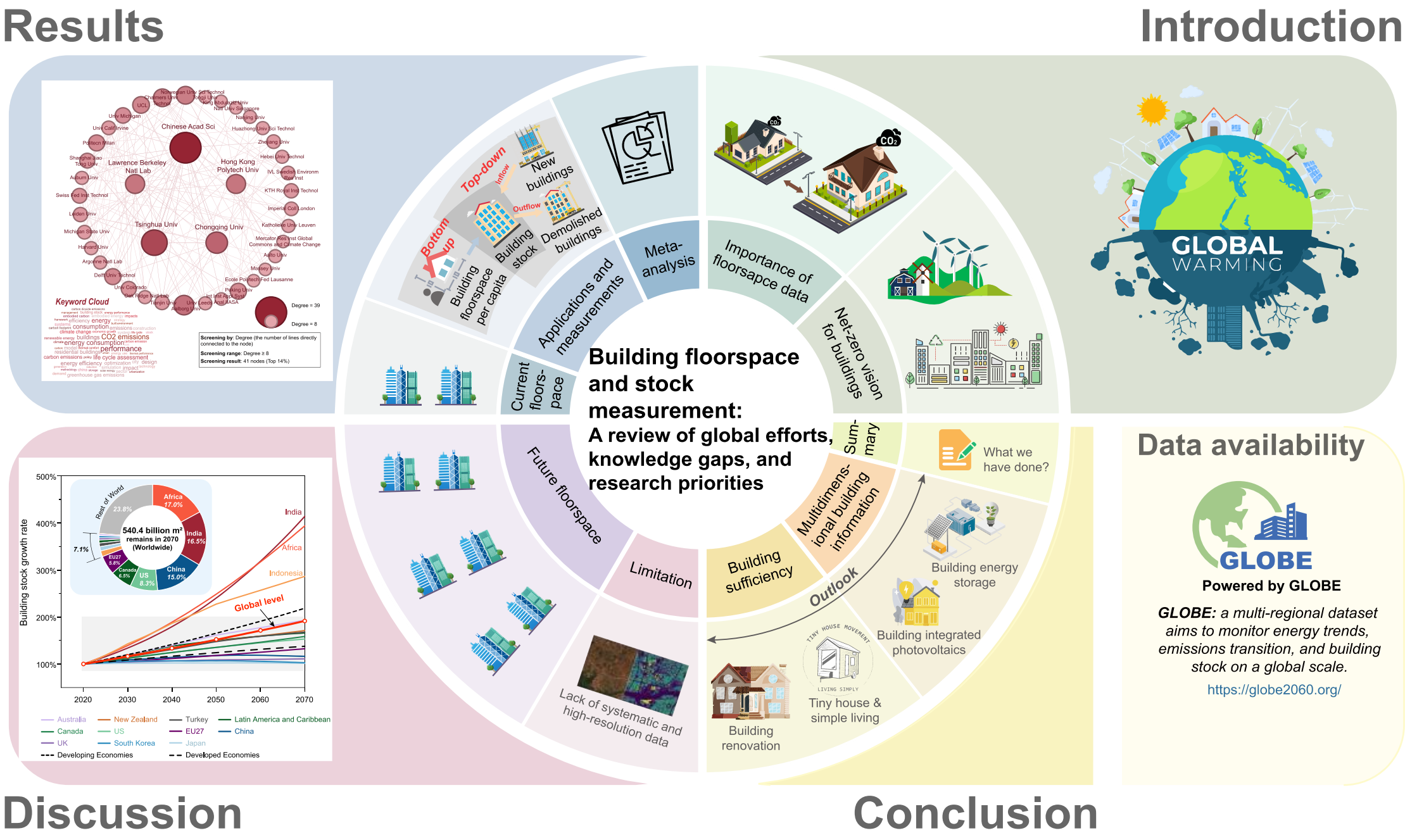

**Graphical abstract.** A review of global efforts, knowledge gaps, and research priorities related to building floorspace and stock measurement, with a focus on energy use and carbon emissions within the building sector.

## HIGHLIGHTS

- We review 2,628 papers on building floorspace and stock, focusing on energy and emissions.
- Floorspace measurement approaches consist of top-down, bottom-up, and hybrid methods.
- By 2070, the global building stock is projected to be 1.87 times its 2022 level (~540 billion $m^2$).
- High-resolution floorspace imagery is essential for advancing low-carbon progress in buildings.
- Building sufficiency is a critical strategy for accelerating decarbonization in the building sector.

## eTOC

Ma et al. analyzed 2,628 papers from the past three decades on global building floorspace measurements. The review underscores measurement challenges and highlights significant floorspace growth, particularly in emerging economies, while calling for a high-resolution floorspace imagery database and advocating sustainable strategies such as building sufficiency to advance net-zero ambitions in the building sector.

## SUMMARY

Despite a substantial body of research—evidenced by our analysis of 2,628 peer-reviewed papers—global building floorspace data remain fragmented, inconsistent, and methodologically diverse. The lack of high-quality and openly accessible datasets poses major challenges to accurately assessing building carbon neutrality. This review focuses on global building floorspace, especially its nexus with energy and emissions. The key research areas include energy modeling, emissions analysis, building retrofits, and life cycle assessments. Each measurement approach—top-down, bottom-up, and hybrid—has its own limitations: top-down methods provide broad estimates but low accuracy, whereas bottom-up approaches are more precise but data intensive. Our simulations reveal a surge in floorspace growth across emerging economies—most notably in India, Indonesia, and Africa—with India's per capita floorspace projected to triple by 2070. We emphasize the need for a high-resolution global floorspace imagery database to compare energy efficiency, track decarbonization progress, and assess renovation impacts while promoting building sufficiency and accelerating the transition to net-zero building systems.

## BROADER CONTEXT

The International Energy Agency announced that the building sector, which is responsible for nearly 40% of global energy consumption and carbon emissions, now faces a significant challenge: the lack of comprehensive data on building floorspace. This data gap hampers accurate measurement of carbon intensity and limits the ability to assess decarbonization strategies throughout the building life cycle. Global efforts to reduce building-related emissions are constrained by fragmented data and methodological limitations. The complexity of quantifying floorspace across diverse building typologies and regional contexts further complicates carbon accounting and mitigation comparisons. This study examines the influence of floorspace on building energy consumption and carbon emissions across all life cycle stages, categorizing existing floorspace measurement approaches and highlighting their strengths and limitations. It also assesses the current global status of building floorspace and provides future projections of the building stock. This study highlights the urgent need for a high-resolution global floorspace imagery database, which would enable more accurate comparisons of energy efficiency, decarbonization progress, and renovation impacts. Additionally, it highlights the importance of building sufficiency—designing and operating buildings to minimize environmental impacts by prioritizing efficiency, downsizing and simple living, optimal space utilization, and reduced resource consumption—as a key strategy for promoting low-carbon, sustainable development in the building sector.

## INTRODUCTION

### Background

Buildings contribute more than 37% of global energy-related carbon dioxide ($CO_2$) emissions,[1] with record-high emissions of more than 13 gigatons of $CO_2$ observed in 2023.[2] These emissions make the building sector a critical focus in global decarbonization efforts. Given the substantial mitigation potential offered by cost-effective and commercially available solutions, the building sector is poised to play a crucial role in achieving net-zero targets by midcentury.[3,4] A key element in assessing and implementing decarbonization strategies is the accurate quantification of building carbon emissions, which are determined by both carbon intensity (emissions per unit of floorspace) and total building floorspace. Thus, reliable and detailed data on global building floorspace are essential not only for measuring current emissions accurately but also for planning future decarbonization pathways across the entire building life cycle.

However, such measurement and planning efforts face a major obstacle: the global lack of comprehensive, high-resolution imagery data on building floorspace. In particular, missing or fragmented data complicate the estimation of carbon intensity and impair the development of region-specific decarbonization strategies. While long-term modeling tools such as our Global Building Stock Model (GLOBUS[5]) attempt to simulate future trends by incorporating building turnover and renovation dynamics (see Figure 1), the absence of detailed imagery and statistical datasets remains a major barrier. Given the diversity in building types, construction practices, and data availability across countries, a one-size-fits-all modeling approach is not feasible. Addressing this critical data gap is a prerequisite for unlocking the full decarbonization potential of the global building sector.

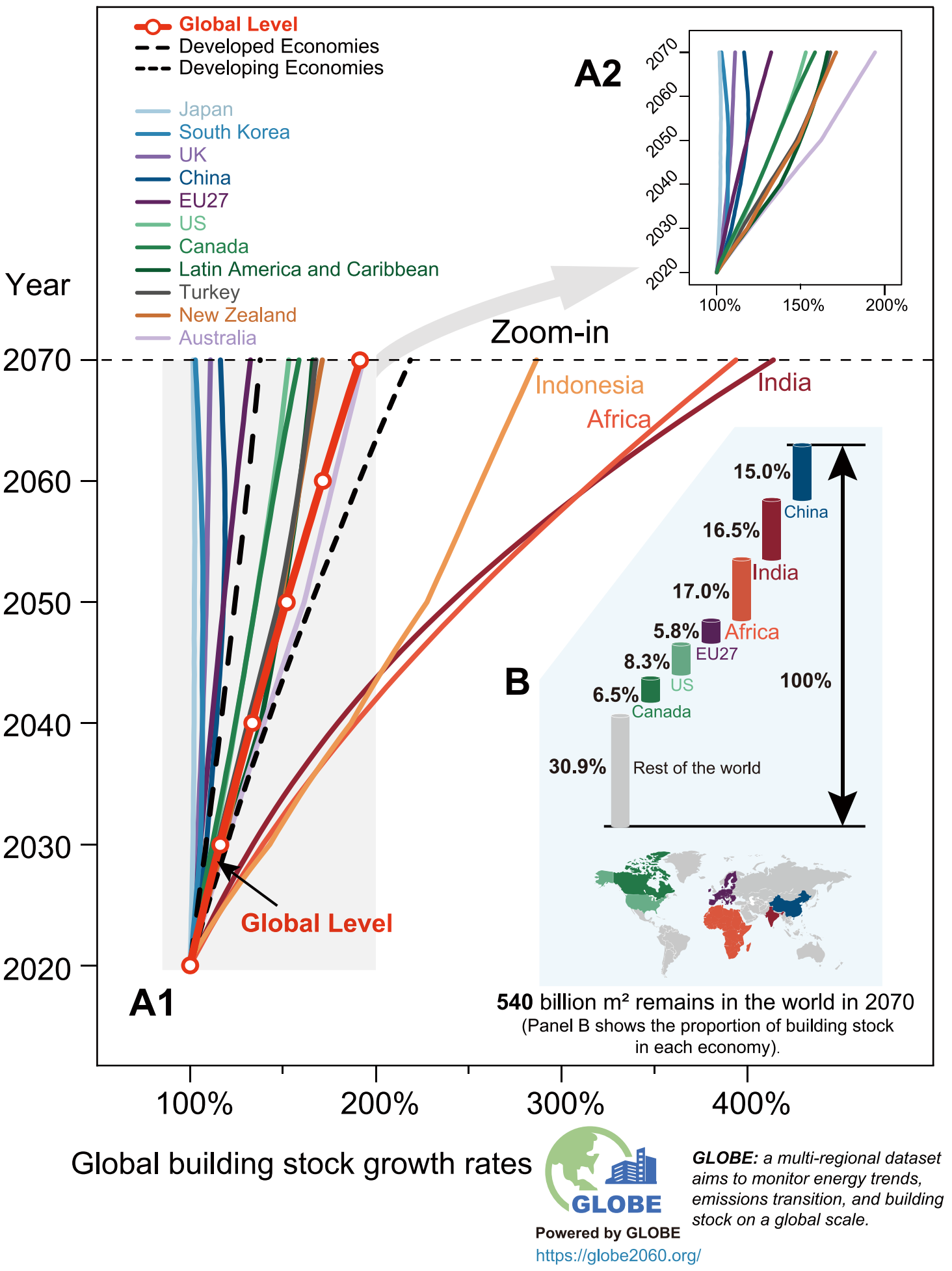

**Figure 1.** Global trends in building floorspace. (A) Global and regional building stock growth rates, 2020-2070; (B) regional stock distribution in 2070. Note: EU27 refers to the 27 countries that are part of the European Union. The classification of developed and developing economies is based on the Human Development Index defined by the United Nations Development Programme (source: *https://hdr.undp.org/data-center/human-development-index#/indicies/HDI*).

### Literature review and research gap

As summarized in Table 1,† in recent years, there has been growing interest in measuring building floorspace and the building stock to better understand trends in energy consumption[7] and carbon emissions.[8] However, existing efforts remain fragmented across institutions, regions, and methodologies, with limited coordination or standardization. Most studies focus on individual countries or case-specific applications,[9-11] often overlooking the broader global context. To address data limitations, various modeling frameworks have been developed, including bottom-up physics-based models[12] and top-down stock turnover models.[13] However, these approaches are frequently constrained by inconsistent or incomplete datasets. Moreover, analyses of the building stock often rely on assumptions or proxy indicators, introducing substantial uncertainty, particularly in cross-country comparisons and long-term projections.[14]

† Note: Table 1 summarizes the top 10 most-cited review articles from our retrieved literature in Table S2, titled "Table S2. Summary of studies on energy and emissions related to building floorspace and stock." The formatting style of Table 1 is adapted from Fazli, et al.[6]

Although an increasing number of studies are linking building stock data with energy and emissions assessments,[15,16] a comprehensive and systematic understanding of global research efforts is still lacking. Key questions remain unanswered: Which institutions and countries are leading research in this field? What methodologies dominate current practices? And how does the growth of floorspace align with building emission trajectories? **These knowledge gaps hinder the comparability of results across regions and limit the accuracy of modeling efforts. Addressing these gaps is essential for advancing data-driven, policy-relevant research on building decarbonization.**

**Objectives and novelty**

To support the development of a global high-resolution imagery database that tracks changes in building floorspace and the building stock, **this comprehensive review aims to** address critical data gaps in current research. Through a systematic bibliometric analysis of studies on floorspace and stock measurements, we examine global efforts, identify knowledge gaps, and highlight research priorities, particularly those at the intersection of energy, emissions, and building floorspace/stock. In this review, we raise three key questions.

- What is the distribution of the retrieved articles among countries, institutions, journals, and keywords?
- What is the current status of applications and measurements of building floorspace and the building stock?
- What are the current and future trends of building floorspace and the building stock in

different economies?

**To address the three questions posed above, we review 2,628 peer-reviewed papers indexed by the Web of Science (WOS) on the topic of building floorspace and the building stock worldwide, with a specific focus on the intersection of energy/emissions and floorspace/stock.** The results of our meta-analysis provide insights for answering the first question, offering a comprehensive overview of the leading countries, institutions, journals, and keywords contributing to this field. For the second question, we further examine the applications and measurements of building floorspace, focusing on two key areas: summarizing the major applications of building floorspace and the building stock and outlining the main approaches used to measure building floorspace. This analysis identifies current practices and highlights gaps in the methodologies employed. Finally, in response to the third question, we conduct an analysis of the global status of building floorspace, estimating future global building stock trends. This analysis provides a broader understanding of the future direction of building stock growth and its potential impact on energy use and carbon emissions. The detailed steps of our bibliometric analysis are provided in the Experimental Procedures section of this paper.

**As the most significant novelty of this study, we present the first and largest systematic review of global building floorspace and stock measurement, highlighting key trends, gaps, and opportunities for improving data accuracy and consistency across regions and methodologies.** Current global efforts to decarbonize the building sector are hampered by fragmented data and methodological limitations. The complexity of measuring floorspace across various building typologies and regional contexts exacerbates the challenge of carbon monitoring and mitigation comparisons. In this review, we examine the influence of floorspace on energy and emissions throughout all building life cycle stages. We categorize existing approaches for measuring floorspace, highlighting their strengths and limitations, and we provide an assessment of the current global status of building floorspace, along with future projections of the building stock. Our findings emphasize the critical need for a comprehensive global floorspace database, which would enable more accurate comparisons of energy efficiency, decarbonization progress, and the impacts of building renovations. Additionally, we highlight the importance

of building sufficiency—beyond just energy efficiency or technological improvements—as a foundational strategy for a low-carbon, sustainable built environment. This study lays the groundwork for future research and policy frameworks aimed at accelerating global decarbonization in the building sector.

**Table 1.** Summary of the top 10 most-cited review articles from our retrieved literature on building floorspace and the building stock in relation to energy use and carbon emissions. Note: Citation data are based on the Essential Science Indicators (ESI) and the WOS Core Collection, accessed on 03/15/2025.

| Reference | Year | Focus | Problems, challenges, and limitations or gaps of science | Recommendations or findings | ESI Highly Cited Paper | Times cited (WOS) |
|---|---|---|---|---|---|---|
| 7 | 2018 | Data-driven prediction of building energy consumption | • Limited transferability of data-driven models across regions and building types<br>• Lack of interpretability and transparency in model parameters | • Enhancing prediction accuracy through hybrid machine learning approaches<br>• Improving model generalizability with innovative architectural designs<br>• Leveraging big energy data to support behavioral energy efficiency strategies | √ | 1145 |
| 8 | 2021 | Trends and drivers of greenhouse gas emissions by sector (1990-2018) | • Ongoing emissions growth fueled by infrastructure expansion and consumption-driven lifestyles<br>• Insufficient strategies to enable rapid technological transitions and sustainable economic pathways<br>• Limited research addressing the phase-out of unsustainable practices and legacy technologies | • Employing targeted decarbonization strategies tailored to specific sectoral needs<br>• Leveraging policy-driven incentives to accelerate emission reductions<br>• Optimizing energy transition pathways to enhance overall efficiency | √ | 585 |
| 9 | 2016 | City-integrated renewable energy solutions for urban | • High costs associated with low-carbon infrastructure investments<br>• Limited adaptability of power grids to the | • Optimized infrastructure planning to support renewable energy integration<br>• Advancement of smart grids, energy storage, | √ | 440 |

| Reference | Year | Focus | Problems, challenges, and limitations or gaps of science | Recommendations or findings | ESI Highly Cited Paper | Times cited (WOS) |
|---|---|---|---|---|---|---|
| | | sustainability | volatility of renewable energy sources<br>• Significant influence of occupant behavior on energy consumption | and integrated energy systems<br>• Promotion of interdisciplinary research to foster sustainable urban development | | |
| 10 | 2018 | Green roofs and facades | • High maintenance and investment costs, along with installation complexities<br>• Energy performance of greenery systems varies significantly by climate and region<br>• Energy savings are contingent on effective and complex moisture management | • Greenery systems significantly reduce energy demand and mitigate heat gain<br>• Energy and thermal performance vary based on plant species and coverage density<br>• Green roofs contribute to carbon sequestration and enhance climate resilience | √ | 350 |
| 11 | 2017 | Recent developments, emerging challenges, and future research directions across the building life cycle | • Lack of standardization in life cycle inventory data collection<br>• Significant discrepancies between calculated and actual environmental impacts<br>• Inconsistent system boundary definitions and challenges in life cycle assessment (LCA) comparisons | • Adopting standardized LCA methodologies to enhance data reliability<br>• Applying dynamic LCA approaches to capture long-term environmental impacts<br>• Improving the integration of LCA into building certification frameworks | √ | 349 |
| 12 | 2017 | Modeling | • Significant variability in urban energy | • Integrating geospatial data to enhance the | | 199 |

| Reference | Year | Focus | Problems, challenges, and limitations or gaps of science | Recommendations or findings | ESI Highly Cited Paper | Times cited (WOS) |
|---|---|---|---|---|---|---|
| | | approaches for urban building energy use | models and their applications<br>• Limited reliability stemming from challenges in model calibration and validation<br>• Insufficient integration of geospatial techniques in urban-scale energy simulations | accuracy of urban energy modeling<br>• Applying advanced calibration techniques to improve model reliability<br>• Optimizing model structures for more scalable and applicable urban energy simulations | | |
| 13 | 2017 | Advancing resource efficiency in the built environment | • Limited integration of circular economy principles in building design and construction<br>• High reliance on linear material flows in urban development<br>• Insufficient research on building stock management and life-cycle-oriented urban planning | • Applying circular economy principles to enhance resource efficiency in the built environment<br>• Optimizing material use to minimize embodied carbon and construction waste<br>• Implementing advanced urban planning strategies to improve building stock management | | 127 |
| 14 | 2022 | Machine learning, deep learning, and statistical analysis for predicting building energy | • High computational resource demands for model training and deployment<br>• Increased risk of overfitting in data-driven forecasting approaches<br>• Lack of standardization in feature | • Employing hybrid modeling approaches that combine physics-based and data-driven methods<br>• Optimizing feature selection and hyperparameter tuning to enhance model robustness | √ | 118 |

| Reference | Year | Focus | Problems, challenges, and limitations or gaps of science | Recommendations or findings | ESI Highly Cited Paper | Times cited (WOS) |
|---|---|---|---|---|---|---|
| | | consumption | selection, data preprocessing, and hyperparameter optimization | • Adopting standardized evaluation metrics for consistent and reliable model comparison | | |
| 15 | 2017 | Urban energy planning for sustainable development | • Insufficient integration between spatial planning and energy systems<br>• Complexity in identifying and implementing optimal urban energy strategies<br>• Limited stakeholder participation in energy planning and decision-making | • Implementing structured urban energy planning frameworks to support informed decision-making<br>• Enhancing planning effectiveness through multisector stakeholder engagement<br>• Optimizing spatial-energy integration for more efficient and sustainable resource use | | 86 |
| 16 | 2022 | Existing policy for reducing embodied energy and greenhouse gas emissions of buildings | • Weak enforcement of regulatory measures at the national level<br>• Heavy reliance on voluntary instruments with inconsistent implementation<br>• Limited availability and transparency of data for embodied energy assessments | • Implementing regulated caps to establish mandatory thresholds for embodied energy and emissions<br>• Mandating LCA reporting to enhance transparency and comparability<br>• Expanding data accessibility and standardizing assessment methodologies to support effective policymaking | √ | 52 |

## RESULTS

### Meta-analysis of the retrieved articles

A systematic literature search conducted in March 2025 identified 2,628 peer-reviewed papers focusing on building energy consumption and carbon emissions related to building floorspace after the exclusion of conference proceedings, data papers, and irrelevant fields. The search spanned more than 30 years, from 1992 to 2025. Among these papers, 46 (1.8%) were classified as ESI Highly Cited Papers. According to the WOS, ESI Highly Cited Papers rank in the top 1%—the standard threshold—by citations within their subject area over the most recent 10-year period.[‡] In our sample, the proportion of such papers nearly doubled this benchmark, underscoring the substantial scholarly attention devoted to research on the interconnections between floorspace, building energy consumption, and carbon emissions.

Figure 2 provides an overview of the retrieved articles. As shown in Figure 2 A, scholars worldwide are actively discussing building energy consumption and carbon emissions at the unit floorspace scale. Notably, scholars from seven countries accounted for 3% or more of this field, including China (22%), the United States (US, 8%), the United Kingdom (UK, 8%), Italy (5%), Spain (4%), Germany (3%), and Australia (3%). Additionally, Figure 2 B highlights the top 41 contributing institutions in this field, comprising 5 core institutions and 36 additional ones. Notably, the 36 institutions were arranged in a circular pattern within the figure. The top five institutions with the most complex collaborative relationships were the Chinese Academy of Sciences, Tsinghua University, Chongqing University, Hong Kong Polytechnic University, and Lawrence Berkeley National Laboratory.

Furthermore, Figure 2 C displays the 27 journals with the highest concentration of relevant papers, with each containing more than 20 retrieved articles and collectively accounting for approximately 70% of all retrieved articles. The five journals with the most retrieved articles were *Energy and Buildings* (262 papers), *Sustainability* (202 papers), *Journal of Cleaner Production* (175 papers), *Energies* (148 papers), and *Applied Energy* (111 papers). Additionally, Figure 2 D presents a keyword cloud derived from the retrieved articles, with the font size and color shading indicating the frequency of use of each keyword. The most popular keywords in this field include “building”, “stock”, “energy”, “consumption”, and “efficiency”.

---

‡ As for the update of the ESI dataset, we referred to the official Clarivate update pages (https://esi.help.clarivate.com/Content/dataset-updates.htm; https://esi.help.clarivate.com/Content/overview.htm). The most recent update was released on March 13, 2025, covering ESI data from 2014 through December 2024. To ensure our bibliometric analysis reflects the most up-to-date dataset, we accessed and retrieved the WOS and ESI data on March 15, 2025, for use in this study.

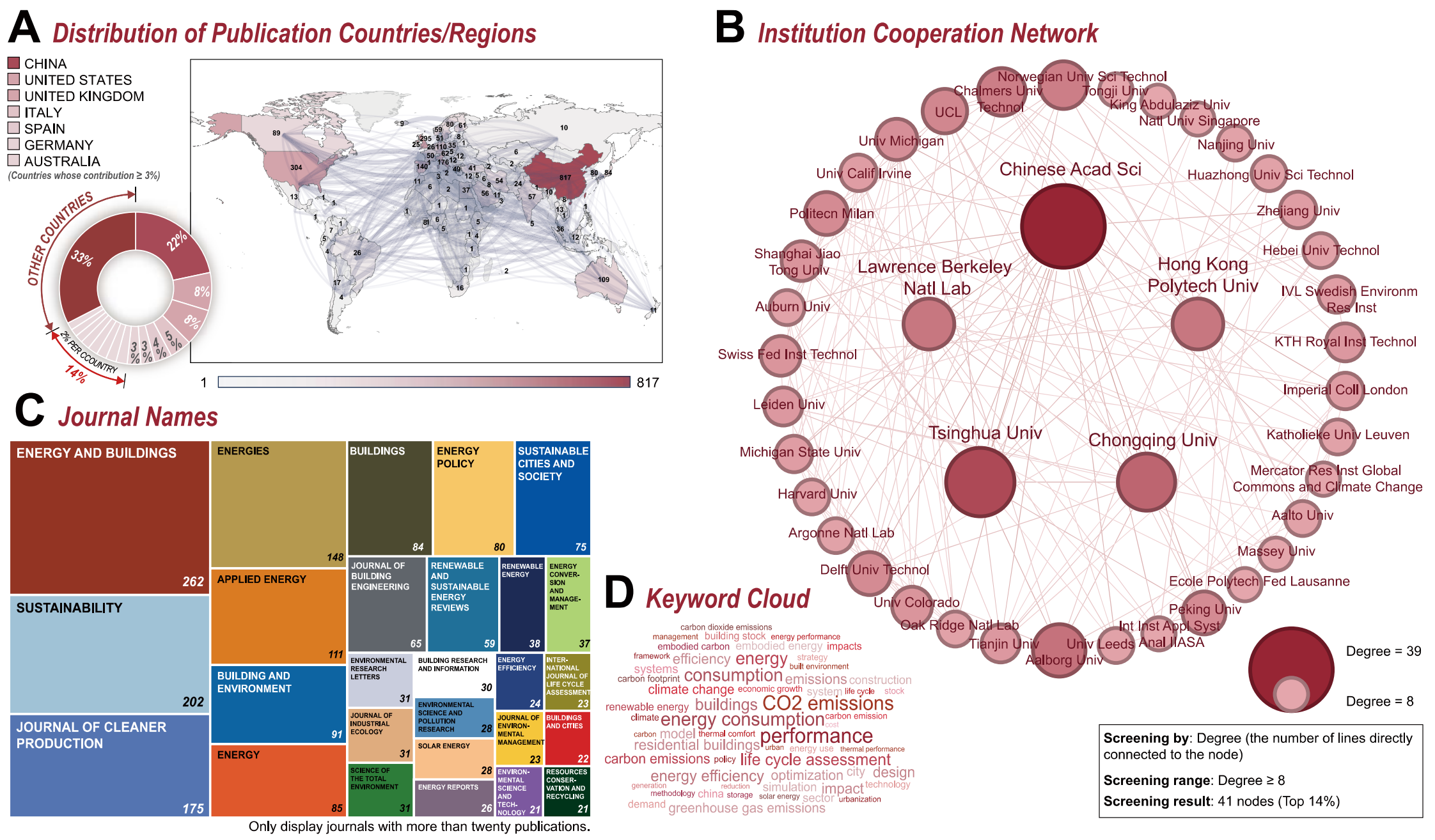

**Figure 2.** Meta-analysis of the retrieved articles. (A) Distribution of publication countries/regions; (B) institution cooperation network; (C) journals with more than ten publications; and (D) the keyword cloud.

To further increase the precision and relevance of the search results, we implemented an additional filtering mechanism using “Keywords Plus” within the WOS and subsequently conducted an in-depth analysis of these retrieved articles alongside the previously identified ESI Highly Cited Papers. After manually screening and excluding low-relevance papers, 114 articles focused on building floorspace remained, 34 of which were ESI Highly Cited Papers. This concentration of ESI Highly Cited Papers further underscores the importance of building floorspace-related research. The content of these 114 articles on the applications and measurement of building floorspace is detailed below. Additionally, the search keyword settings, process, and specific results are provided in Sections 1-3 of the Supplemental Information.

**Applications and measurement of building floorspace**

Building floorspace has distinct implications across all stages of the building life cycle. During the pre-construction and construction stages, energy use and carbon emission intensities are quantified per unit of newly constructed floorspace. In the operation stage, these intensities are typically measured against the building stock, which is defined as the total floorspace of operational buildings. This measurement reflects the net balance between floorspace inflow (i.e., newly constructed and renovated buildings) and outflow (i.e., demolished buildings). During the demolition stage, the floorspace of buildings that have reached the end of their service life and require demolition serves as the denominator for calculating energy and carbon emission intensities.

Before delving into the specifics of building floorspace quantification, the first step is to understand when building floorspace quantification is required across various building types, stages, and goals and which approaches are used in this quantification. To address these questions, we categorized the retrieved articles involving building floorspace and attempted to respond to the following queries:

- Building Type: What type of building was the research question focused on (e.g., residential or non-residential)?
- Stage: Which stage of the building life cycle did the research address (e.g., pre-construction, construction, operation, demolition, or across all stages)?
- Goal: What were the goals of using building floorspace to measure carbon emissions (e.g., energy conservation, emission reduction, climate protection, or sustainable development)?
- Approach: How was building floorspace quantified (e.g., top-down, bottom-up, or a hybrid approach combining both)?

Figure 3 illustrates the hierarchical structure of building floorspace applications in the retrieved articles, including building types, life cycle stages, topics, methods, and goals. The width of each area represents the number of studies associated with the corresponding application. With respect to building types, the distribution of studies is as follows:

residential buildings (61%), non-residential buildings (5%), and civil buildings (34%).[§] From the perspective of the building life cycle, studies have focused on the following stages: pre-construction (6%), construction (13%), operation (66%), and all stages (15%). It is clear that the majority of studies on energy consumption and carbon emissions related to building floorspace focus on the operation stage of residential buildings. The goals of using building floorspace are as follows: energy conservation (48%), emission reduction (40%), climate protection (4%), and sustainable development (8%).

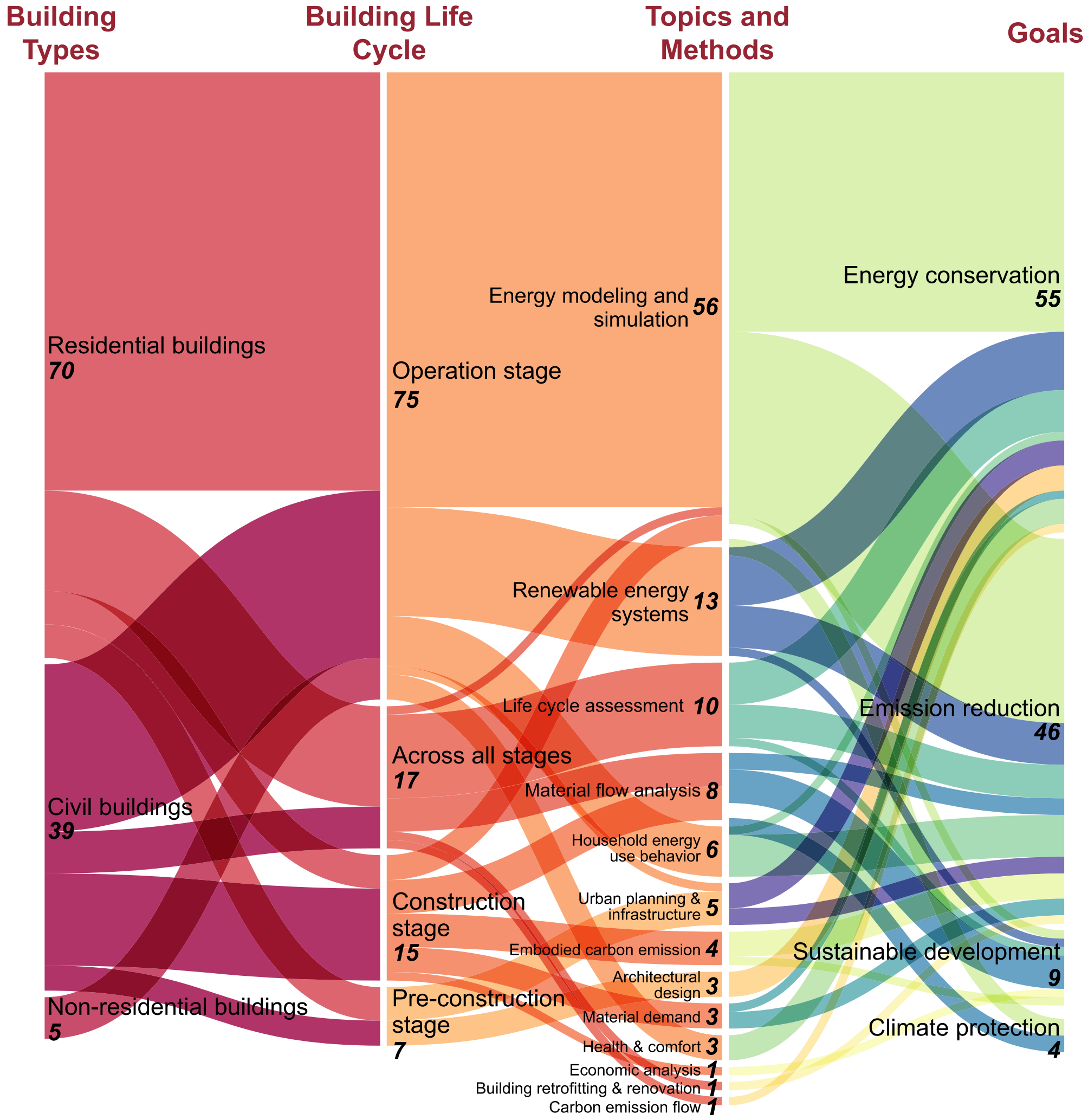

**Figure 3.** The hierarchical structure of building floorspace applications in the retrieved articles, categorized by building types, life cycle stages, research topics and methods, and goals. Note: the width of each area reflects the number of articles associated with each category.

§ Civil buildings include the residential and non-residential buildings.

Among these applications, building floorspace plays a critical role in determining building energy consumption and carbon emissions. While it is primarily used as an intermediate metric in environmental assessments—especially for energy and emissions—its fundamental role in shaping a building's overall environmental impact remains underexplored. Therefore, this review first introduced the applications of floorspace in measuring energy and carbon emissions across various building types and stages, followed by a detailed analysis of the methodologies used to measure floorspace.

**a. Applications of building floorspace featuring carbon intensity**

The retrieved articles on applied building floorspace are categorized into two groups: direct applications (58% of the total) and applications after quantification (42% of the total). The distinction between these two types lies in whether the research includes the quantification of building floorspace. The following section provides a detailed explanation of the direct application of building floorspace.

In the field of energy efficiency assessment and building decarbonization, building floorspace is often used as a fundamental parameter to evaluate energy consumption and emissions. While crucial in these complex studies, floorspace is seldom the primary focus. As a result, some studies have directly utilized existing floorspace data for more detailed analysis of building-related factors, provided that sufficient data are available. This method is especially prevalent in urban planning,[15,17,18] infrastructure development,[19-22] and architectural design in the pre-construction phase of the building life cycle.[23] Additionally, the direct use of building floorspace data for research on building energy consumption and carbon emissions is more common during construction, operation, or the whole life cycle of buildings.[24-26] The applications mentioned above are broad, encompassing but not limited to the following areas:

- Embodied decarbonization: Exploring the impact of reduced new construction and improved material efficiency on decarbonization goals.[27-29]
- Energy and emission models: Using various models for simulating energy use in buildings,[30,31] alongside energy forecasting models,[14,32,33] as well as decarbonization models[34] for evaluating carbon peak and neutrality, with building floorspace as a key parameter.[2,35]

- Factors affecting carbon emissions: Analyzing the influence of population,[36] gross domestic product,[37] building floorspace,[38-40] and household conditions[41,42] on building emissions.[43]
- Energy efficiency improvements: Assessing the energy efficiency of building operations through building renovation,[44,45] particularly in terms of space heating[46,47] and cooling technologies,[48] with a focus on energy savings, decarbonization benefits,[49] and cost benefits.[50,51]
- Building-integrated photovoltaics (BIPVs): Evaluating the development of BIPVs in distributed energy systems, influenced by building floorspace.[52,53]

While building floorspace data are readily available for certain regions, comprehensive global datasets disaggregated by building type and geography remain scarce. This widespread lack of floorspace statistics significantly hampers the measurement and evaluation of building energy consumption and carbon emissions at the intensity level, thus limiting the scope and depth of building-related research.

**b. Measurement of building floorspace**

While a subset of the retrieved articles did not directly focus on the statistical analysis of floorspace, these studies nevertheless explored the quantification approaches for floorspace to some degree. Since the accounting of building floorspace is usually integrated with the accounting process of building energy consumption and carbon emissions, existing research on building energy consumption can be divided into three main categories, i.e., top-down, bottom-up, and hybrid models, which are classified by the modeling method;[12,54] or white, black, and gray boxes, which are classified by the degree of transparency.[7,55] Therefore, the quantification of building floorspace can refer to the classification logic used in the classification of building energy consumption research, which is categorized into a top-down approach starting from a macro perspective, a bottom-up approach starting from a micro perspective, and a hybrid approach that combines the top-down and bottom-up approaches.

**Top-down approach:** The top-down approach mainly refers to the building stock turnover model, which combines inflow (newly constructed and renovated buildings)[56] and outflow (demolished buildings)[57,58] buildings at the macro level of a region to count the net stock of regional buildings (see Figure 4). It usually adheres to the entire life cycle of buildings and focuses on analyzing newly constructed buildings and the building stock.[59] Building stock turnover is often linked to the turnover of building materials. Some studies have combined these processes to analyze embodied carbon emissions associated with materials throughout the building life cycle,[60] particularly during the construction stage.[61] Furthermore, the results of the top-down building stock turnover model provide a foundation for bottom-up analyses of the energy consumption of end-use activities in building operations.[62] In addition, some studies have used the number of households to represent the building stock, which was calculated by dividing the population by the average household size. While this method is a top-down approach, it applies only to residential building research.[63]

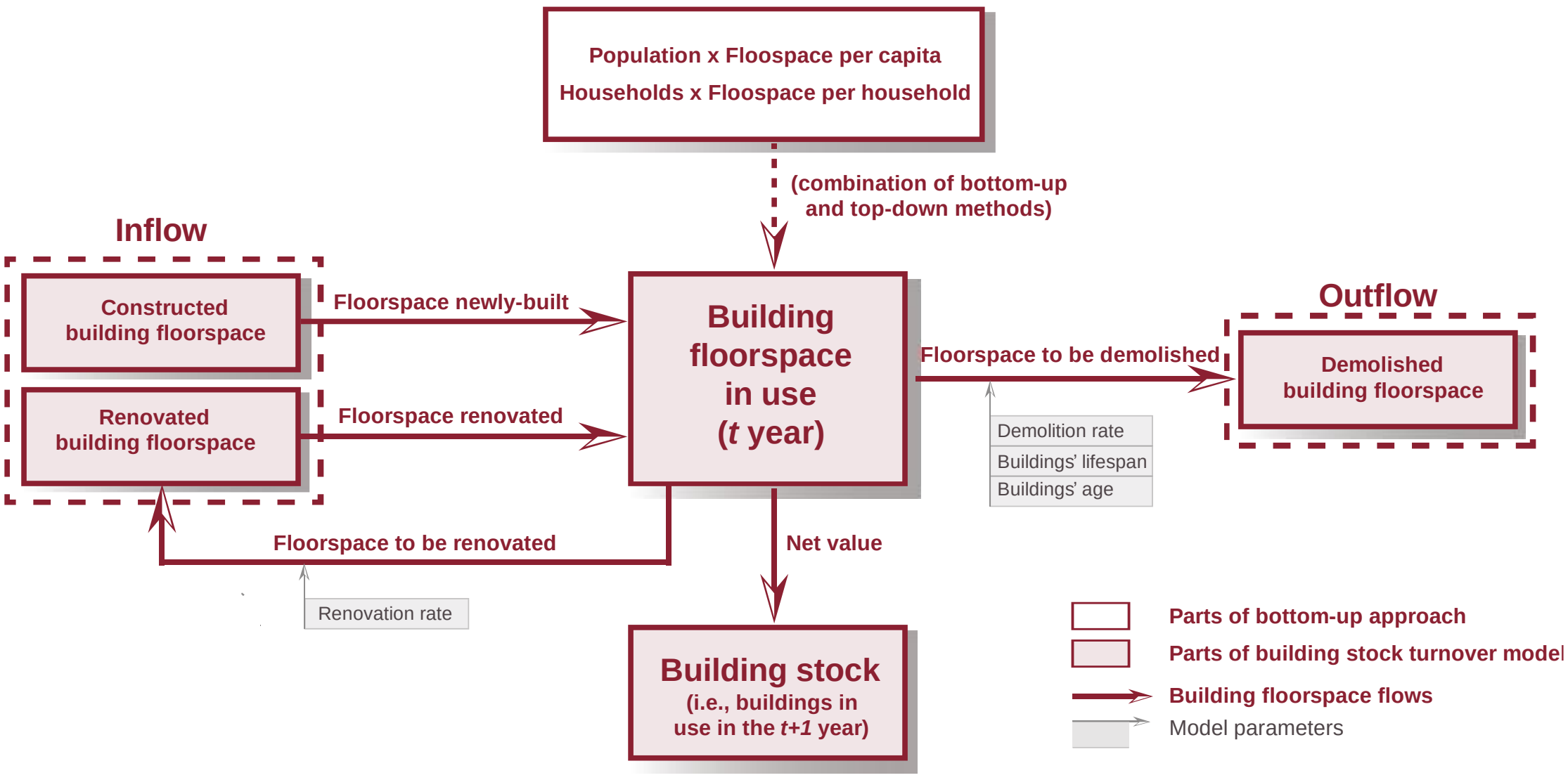

**Figure 4.** Schematic diagram of the building stock turnover analysis process.

Owing to the research paradigm of the top-down approach, its advantages and limitations are obvious. Compared with the bottom-up microdata-based approach, the top-down approach adopts a macro perspective that is applicable to a broader range of research domains. In addition, the top-down building stock turnover model can provide more comprehensive macro data related to building floorspace, covering the floorspace of newly constructed, existing, and demolished buildings, which is conducive to top-down analysis of building energy consumption and embodied and operational carbon emissions at the macro level. However, the building stock turnover model is based on macro analysis, with the parameters reflecting social averages, which limits its ability to capture spatial differences. For example, the building lifespan and renovation rate involved in the model may show significant regional differences under different climatic and economic conditions. Additionally, the results provided by the top-down approach have limited guiding significance for specific technical improvements to enhance building energy efficiency.

**Bottom-up approach:** The bottom-up approach can be divided into two main categories: the demand-driven and physical modeling approaches. The demand-driven approach considers per capita comfort and environmental quality, estimating the regional building stock by multiplying the regional population by the estimated suitable per capita building floorspace[64] or by multiplying the number of regional households by the estimated suitable building floorspace per household[65] [e.g., Residential Energy Consumption Survey of U.S. Energy Information Administration (EIA)]. Moreover, for commercial buildings, some studies estimated total building floorspace by multiplying the number of service personnel by the building floorspace per employee.[66] The demand-driven bottom-up approach is widely used to predict future carbon emission pathways because of its simplicity, enabling quick estimation of building floorspace data and its impact on carbon emissions.[67,68]

The second type of approach is based on physical modeling and is both technology-oriented and data-driven. Two main forms are used: the first form integrates geographic information systems and remote sensing data to perform physical modeling on a small area,[69,70] such as cities[71-73] or neighborhoods,[74] to estimate the regional building stock.[75-77] The other form classifies buildings according to typical characteristics,[78-80] selects representative examples to build physical models,[81,82] and extrapolates the results on single building floorspace and energy consumption to the regional level,[83,84] enabling analysis of regional building floorspace and energy consumption from point to surface. Both forms of this approach are often used to quantify the energy efficiency improvements,[85] decarbonization benefits,[86] and cost effectiveness[87] resulting from building renovations.[88] In addition to analyzing the regional building stock from point to surface, this physical modeling approach has been applied to the study of specific buildings or specific building types, such as schools.[89,90]

Like the top-down approach, the bottom-up approach also has obvious advantages and disadvantages. As the main advantage of the demand-driven bottom-up approach, the calculation process is simple, and the demand for raw data is relatively small. However, the applicability of this approach largely depends on the availability of key per capita floorspace data. This approach is suitable for national-level studies, allowing separate calculations for residential and non-residential buildings. However, because it relies on per capita floorspace, it is less applicable to studies of specific building types.

The bottom-up physical modeling approach has significant advantages. Compared with the top-down approach, the estimation results of building floorspace are closer to the actual situation and can provide detailed floorspace data, including building size, the door and window layout, envelope data, the roof area, etc. This information is crucial for evaluating the photovoltaic potential of buildings, improving energy efficiency in space heating and cooling, and evaluating the effects of building renovations and equipment upgrades. This demonstrates that the bottom-up physical modeling approach offers greater breadth and depth in application. Furthermore, this approach is more flexible than the top-down approach in estimating the building stock of a specific area or building type from a micro perspective. Although the bottom-up approach has many advantages, its application relies on large and complex basic data, which limits its applicability at the national or global level.

**Hybrid approach with top-down and bottom-up approaches:** The hybrid approach combines the top-down building stock turnover model with the demand-driven bottom-up approach.[91] In applying the top-down building stock turnover model, some studies derived a static building stock by multiplying population size by the per capita building floorspace or the number of households by the average building floorspace per household as the input data of the model[65] (as shown by the dotted line in Figure 4), thus reflecting an integration of the top-down and bottom-up approaches and reflecting a coordination process for top-down data and bottom-up data. In addition, to increase research reliability, some studies classified regional buildings by typology and applied building stock turnover models to each of them,[92] which also reflects the combination of top-down and bottom-up approaches.[93] Hybrid approaches are commonly employed to calculate the potential future building stock,[94] energy consumption,[95] and carbon neutrality pathways under various economic, technological, and emission scenarios.[96] The integration of the demand-driven bottom-up approach has extended the time scale of the building stock turnover model, facilitating the analysis of future renovation potential.[97] In contrast, the physical modeling method encounters greater challenges in forecasting future pathways.

Although bottom-up physical modeling is rarely used to predict floorspace in hybrid approaches, it can improve the accuracy of building stock assessments beyond the capabilities of other methods. With the rapid advancement of artificial intelligence (AI) technology and the decreasing barriers to its use, the difficulty of coordinating and calibrating top-down macro-statistical data with remote sensing data obtained through technology-oriented bottom-up methods (i.e., geographic information systems and remote sensing technologies) can be reduced, thus significantly improving the accuracy of building stock assessments. Specifically, statistical and remote sensing data can be coordinated and integrated through data assimilation, with AI identifying data deviations and machine learning consolidating scattered or incomplete data sources to increase the accuracy of building stock assessment.[98]

Each approach—top-down, bottom-up, or hybrid—has its own distinct advantages and limitations. The choice of an appropriate building floorspace quantification method should be aligned with the specific needs of the research. Despite the widespread use of floorspace data in building-related studies, in-depth investigations into quantification methodologies are still lacking. The development of a high-resolution and comprehensive building floorspace imagery database is urgently needed to advance this area of research.

**Current global building floorspace status**

While the preceding analysis underscores the pivotal role of floorspace data, global datasets remain sparse, especially continuous building floorspace data with time series. Among the retrieved articles, less than half addressed the quantification of building floorspace, with time series data on building floorspace being notably scarce. Therefore, we expanded the scope of our search, referring to the research results of authoritative institutions (e.g., the United Nations Environment Programme,[99] the Global Alliance for Buildings and Construction,[100] and the International Energy Agency[101,102]), high-quality articles published by native scholars from different countries,[103-105] and manually screened and summarized building floorspace data with reference values. The summarized regions include eight developed economies, namely, the US, Canada, Japan, South Korea, the UK, New Zealand, the EU27 and Australia, and six emerging economies, namely, China, India, Africa, Turkey, Indonesia, and Latin America and the Caribbean (LAC). Given comparable conditions, we analyzed global trends in per capita floorspace, as shown in Figure 5, which depicts the per capita floorspace of residential and non-residential buildings across 14 economies from 2000-2022.

For the residential buildings shown in Figure 5 A, the per capita floorspace in developed economies substantially exceeded that of emerging economies, ranging from 35.8 square meters[**] ($m^2$, South Korea) to 62.4 $m^2$ (US) in 2022. However, in most emerging economies, the per capita residential floorspace was well below 30 $m^2$, with Indonesia having the lowest value at approximately 11.3 $m^2$. In China and Turkey, although they are emerging economies, in 2022, their per capita residential floorspace reached approximately 39.5 and 36.7 $m^2$, respectively. Additionally, the per capita residential floorspace in the US varied greatly, which to some extent is determined by the statistical method of residential buildings in the US. The EIA conducts a Residential Energy Consumption Survey approximately every five years; it is a sampling survey that counts the residential building floorspace in that year. This introduces significant randomness in sample selection, which greatly impacts the final results. Additionally, the per capita residential floorspace in other regions generally tends to increase.

[**] One square meter is equal to 10.764 square feet.

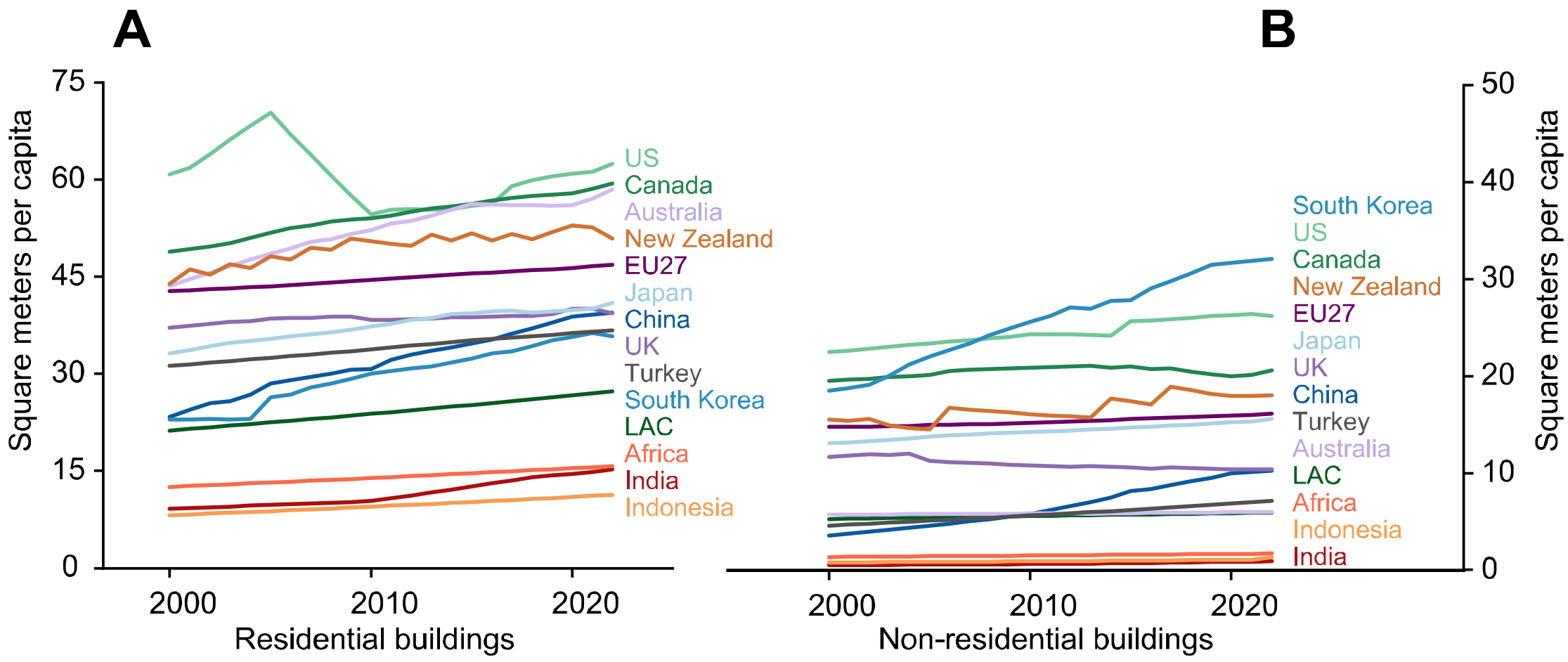

**Figure 5.** Trends in per capita floorspace for (A) residential and (B) non-residential buildings worldwide, 2000-2022.

As shown in Figure 5 B, the per capita non-residential floorspace in developed economies was also generally greater than that in emerging economies. Compared with other developed economies, South Korea's per capita non-residential floorspace grew rapidly and reached a maximum of 32.1 $m^2$ in 2022, driven by a rapid decline in population after reaching a peak. Other developed economies with an increasing trend in non-residential floorspace per capita were distributed mainly between 15.6 and 26.2 $m^2$ in 2022. Owing to Australia's small per capita non-residential floorspace base and slow growth between 2000 and 2022, the per capita non-residential floorspace in 2022 was only 6.0 $m^2$. In addition, the UK's long-term rapid population growth caused by net immigration has led to a downward trend in its per capita non-residential floorspace, which was only 10.4 $m^2$ in 2022. With respect to emerging economies, only China, Turkey, and LAC had a slightly higher per capita non-residential floorspace than Australia did in 2022, at approximately 10.3, 7.2, and 6.0 $m^2$, respectively. Other emerging economies were significantly less common, with India at approximately 0.9 $m^2$, Africa at approximately 1.8 $m^2$, and Indonesia at approximately 1.4 $m^2$.  We believe that the differences in non-residential floorspace per capita among economies can largely be attributed to the unequal development of each economy's service industry and the varying population sizes across these economies.

## DISCUSSION

In this section, we focused on several key implications of this review article, including future global building stock estimations, as well as the limitations and future outlook of this work.

### Future global building stock estimations

Figure 5 illustrates the global trends in per capita floorspace from 2000-2022. On this basis, we further collected and collated the projected development of global floorspace under the business-as-usual (BAU) scenario. Given comparable conditions, we conducted a preliminary analysis of the potential growth in per capita residential and non-residential building floorspace for 14 economies from 2022-2070 under the BAU scenario (see Figure 6). This analysis is based on representative studies by authoritative institutions (e.g., IEA[106-108], EIA[109]) and native scholars.[110-113] The detailed scenario setting and assumptions are provided in Section 4 of the Supplemental Information. With respect to the growth rates for per capita residential building floorspace presented in Figure 6 A, emerging economies generally outpace developed economies. For example, emerging economies were projected to achieve an average growth of 61.4% by 2070 compared with 2022, whereas developed economies may grow by only 35.3%. India, the fastest-growing region, was expected to reach approximately 3.2 times its 2022 per capita residential floorspace by 2070. Additionally, Indonesia and Africa were expected to be significant drivers of global residential building floorspace growth, with projected increases of approximately 80.1% and 63.8%, respectively, by 2070. Other emerging economies, such as China, Turkey, and LAC, were expected to experience growth rates similar to those of most developed economies, ranging between 26.2% and 48.0%, close to the global average growth rate of 42.3%. Conversely, the UK was projected to experience minimal growth in per capita residential floorspace, increasing by only approximately 5% by 2070. This modest rise is likely due to rapid population growth, which nearly matches the rate of residential floorspace expansion.

The dotted error bands in Figure 6 A represent other possible growth rate ranges for per capita residential floorspace at key time points (e.g., 2030, 2040, 2050, and 2060) under the BAU scenario, accounting for uncertainties. Berrill, et al.[114,115] proposed that the per capita residential floorspace in the US may grow along a faster path and may be 16.3% higher than the EIA's forecast by 2060. Moreover, Cabrera Serrenho, et al. [116] noted that the per capita residential floorspace in the UK in 2050 may fluctuate between -34.1% and 25.2% of the level determined by Drewniok, et al. [117] (shown by the purple solid line). In addition, the results of Hong, et al. [118] show that China's per capita residential building floorspace in 2050 may fluctuate between -28.9% and 34.4% on the basis of the blue solid line. To synthesize the data, we utilized the most recent research findings as the basis for the solid trend line. While the reference data for uncertainty bands are relatively dated, these historical results are retained to provide comparative insights, given both the substantial uncertainties in floorspace quantification and the limited data availability.

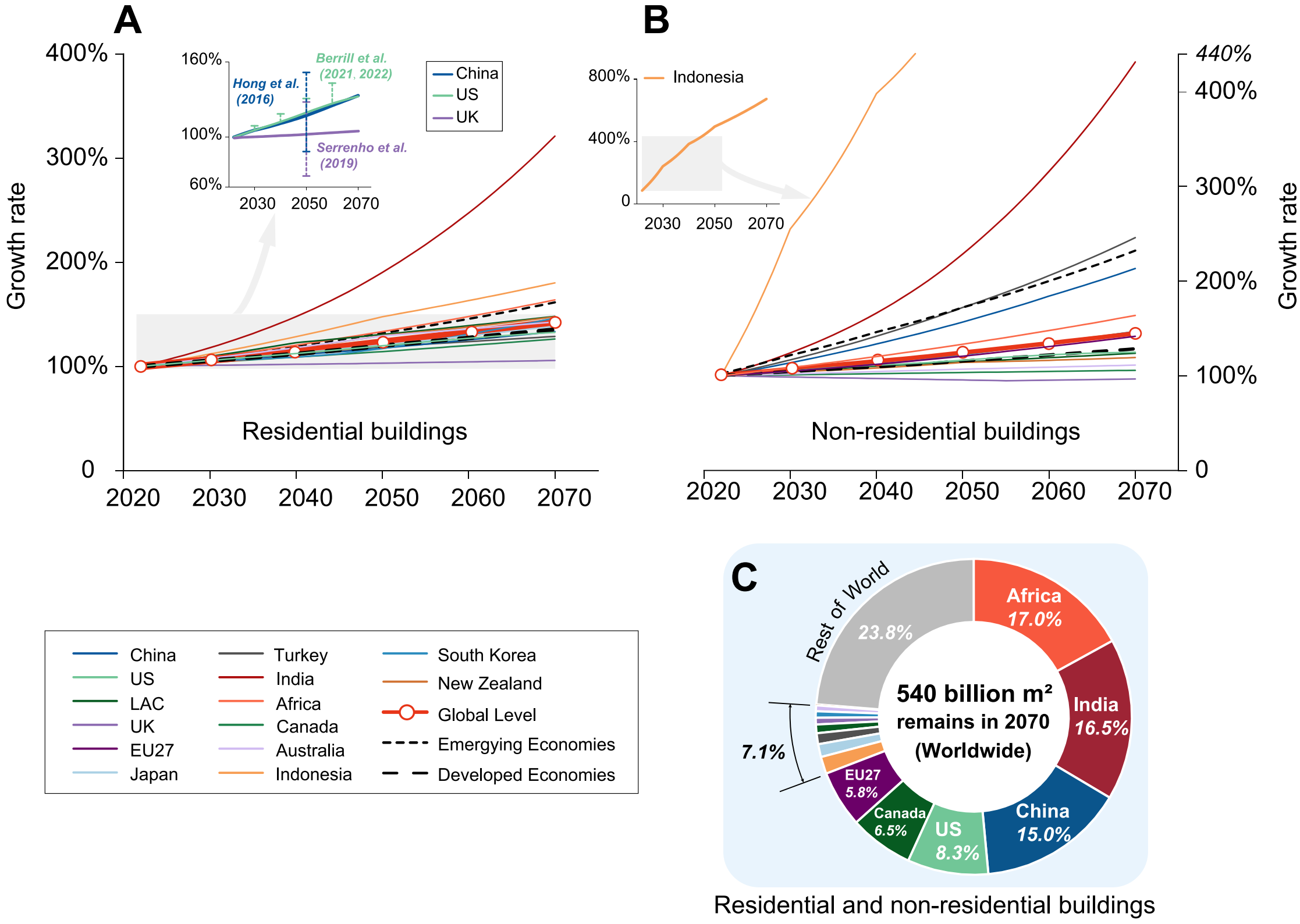

**Figure 6.** Growth trends in per capita floorspace for (A) residential and (B) non-residential buildings worldwide under the BAU scenario, 2022-2070; (C) the global building stock level in 2070.

Figure 6 B shows that under the BAU scenario, growth in per capita non-residential floorspace parallels that in residential floorspace, with emerging economies generally experiencing faster growth than developed economies. For example, by 2070, emerging economies could grow more than twice as much as they did in 2022, whereas developed economies may grow by only 27.9%. Notably, the per capita non-residential floorspace in Indonesia was projected to grow significantly, reaching 6.9 times its 2022 level by 2070. This sharp increase is attributed to both rapid expansion and Indonesia's initially low baseline, with a per capita non-residential floorspace of approximately 1.4 $m^2$ in 2022. While India's growth rate in per capita non-residential floorspace is not as fast as that of Indonesia, it is still much higher than that of most other economies, reaching 4.3 times its 2022 level by 2070. However, India holds the greatest potential for total non-residential floorspace growth, with its population projected to reach 1.69 billion by 2070, which is 5.3 times greater than that of Indonesia.

Although the per capita non-residential building floorspace in China and Turkey exceeded that of Australia in 2022, it remains significantly lower than that in other developed economies. As shown in Figure 6 B, both China and Turkey have considerable growth potential, with projected increases of 113.2% and 145.9%, respectively, by 2070. Among emerging economies, Africa and LAC show relatively modest growth in per capita non-residential floorspace, with projected increases of only 63.8% and 23.8%, respectively, by 2070. In developed economies (excluding the UK), the per capita non-residential building floorspace was projected to grow slowly, with increases ranging from 5.8% to 46.9% by 2070, with most falling below the global average growth rate of 44.2%. In the UK, where population growth was expected to outpace non-residential building floorspace expansion, the per capita non-residential floorspace may decline by 3.4% by 2070. Additionally, owing to the limited availability of data on non-residential floorspace, further discussion of the uncertainty in per capita non-residential floorspace across economies is difficult. In summary, as illustrated in Figure 6 C, by 2070, the global building stock was expected to be approximately 1.87 times its 2022 level (approximately 540 billion $m^2$), posing a significant challenge for sustainable development in habitats and the built environment.

**Limitations and future outlook**

On the basis of the findings in the RESULTS section, we identified three key limitations and their associated solutions, as outlined below:

a. **Systematic and high-resolution measurements of building floorspace data are urgently needed.** While some studies have explored methods for quantifying building floorspace, in-depth research in this area remains insufficient. Accurate acquisition of building floorspace data and the comprehensive establishment of a global building floorspace database are still critical issues that need to be addressed. A comparison of top-down, bottom-up, and hybrid approaches for building floorspace measurement reveals that the top-down approach offers comprehensive system coverage but struggles with accuracy. Additionally, the bottom-up physical modeling approach can provide precise data but faces significant challenges in creating large-scale databases. Finally, the hybrid approach, which typically combines building stock turnover models with demand-driven bottom-up methods, also falls short of realizing a precise database.

b. **Focusing on the sufficiency of building floorspace and maximizing building utilization are essential goals.** Importantly, building floorspace serves as the foundation for various energy-consuming activities throughout a building's life cycle. The size of the floorspace fundamentally determines the energy and emissions at each stage of a building's life. Therefore, extending the lifetime of buildings through renovation can help prevent unnecessary reconstruction, thus reducing energy consumption and carbon emissions. Specifically, in the construction stage, leveraging building sufficiency can reduce the upstream and downstream carbon emissions associated with the production, transportation, and use of materials required for new buildings. In the operation stage, maximizing building sufficiency through renovation can lead to energy conservation and emission reduction benefits through improved energy efficiency. Finally, in the demolition stage, utilizing building sufficiency can help avoid unnecessary demolition, thus mitigating the associated energy consumption and emissions. Additionally, more attention should be paid to the development of tiny houses that downsize buildings and promote a simple living style. Overall, prioritizing the sufficiency of building stock is essential for fully utilizing available resources and achieving sustainable development.

c. **Measuring multidimensional building information enhances high-resolution building stock assessments and improves photovoltaic potential evaluations.** In addition to floorspace information, multidimensional building information includes the building footprint, height, and the envelope area.[119] The building footprint and height are crucial for calculating both building floorspace and the envelope area. While the accurate quantification of building floorspace is essential, developing a comprehensive record of the building envelope, including roofs and exterior walls, creates significant analytical possibilities. For example, teams such as Google AI have recently launched the Open Buildings 2.5D Temporal Dataset.[98] This dataset leverages machine learning technologies, such as the teacher–student model, combined with high-resolution satellite imagery from Sentinel-2, to simulate building presence, height, and fractional building counts in the Southern Hemisphere. In the future, AI-driven high-resolution imagery calculations will significantly accelerate the measurement of multidimensional building information, advancing high-resolution building stock assessments and improving photovoltaic potential evaluations. By then, real-time monitoring of changes in the global building stock will also be possible. The building envelope, which serves as the primary physical infrastructure for BIPVs, plays a pivotal role in determining the photovoltaic power generation potential. Enhancing and refining building stock accounting will facilitate the full realization of BIPV potential, drive the widespread adoption of building photovoltaic power systems, and support the transition of buildings from simple energy consumers to distributed energy suppliers within grid-interactive systems.

## CONCLUSION

This study presented a comprehensive analysis of the global literature and data landscape related to building floorspace and the building stock, with a particular focus on their roles in energy use and carbon emissions assessments. Through a systematic review of peer-reviewed articles and forward-looking projections under a BAU scenario, we highlighted the current state, methodological approaches, and emerging directions in the field of building floorspace and stock measurement. The main findings and their implications are summarized below.

**Key findings**

- **Global research on building floorspace and stock measurements has expanded significantly in recent years, with China and Western countries emerging as the leading contributors.** Among the 2,628 peer-reviewed articles analyzed from the WOS, spanning the past three decades (1992–2025), China, the US, and the UK accounted for more than 60% of all publications. Notably, three of the five most active research institutions in this field—the Chinese Academy of Sciences, Chongqing University, and Tsinghua University—are based in mainland China, underscoring the region's prominent role in advancing knowledge on the building stock and its environmental implications. Other key institutions include Lawrence Berkeley National Laboratory (US) and Hong Kong Polytechnic University.
- **High-impact research on building floorspace has focused primarily on residential operations and spans the entire building life cycle.** From the 2,628 papers screened, 114 highly relevant studies were identified through targeted filtering and manual review. Notably, 26% of these studies were classified as highly cited in the ESI database, highlighting the increasing academic interest in this field. While the majority of these studies (approximately 60%) concentrated on the operational stage of residential buildings, they also addressed floorspace applications throughout the construction and use phases. The key research areas include material flow analysis, embodied carbon emissions, energy modeling and simulation, renewable energy systems, health and comfort, household energy use behavior, and building retrofitting and renovation, demonstrating the diverse role of floorspace in energy and emissions assessments.
- **Three main categories of building floorspace estimation approaches were identified: top-down, bottom-up, and hybrid methods.** The top-down approach, which is primarily based on building stock turnover models, estimates floorspace at a larger scale from a macro perspective, but it tends to have slightly lower accuracy. The bottom-up approach includes demand-driven and physical modeling methods. The demand-driven approach is relatively simple and relies mainly on population size and the per capita floorspace. In contrast, the physical modeling approach can achieve high accuracy, but it is time consuming and requires substantial amounts of high-quality data for training and deployment. The hybrid approach combines elements of both top-down and bottom-up methods, typically integrating the building stock turnover model with the demand-driven approach. However, it still faces challenges in accurately quantifying building floorspace.

- **Global trends in per capita building floorspace show significant disparities between developed and emerging economies, with emerging economies expected to experience rapid growth.** Since the turn of the millennium, the global per capita floorspace for both residential and non-residential buildings has increased gradually, with developed economies being consistently higher than emerging economies. However, our projections indicated that emerging economies would outpace global averages in floorspace growth from 2023 to 2070. Specifically, per capita floorspace in these economies was expected to increase by 129.9% by 2070, compared with only 44.2% globally. The key emerging economies driving this growth include India, Indonesia, and parts of Africa, with India projected to see its per capita residential floorspace more than triple by 2070, while Indonesia's non-residential floorspace was expected to grow nearly sevenfold.

**Policy and research implications**

Through our analysis of applications, measurement approaches, and data simulations related to building floorspace and the building stock, we emphasized that reliable floorspace data are fundamental for building science research. Accurate measurement enables cross-country comparisons of energy conservation and emission reduction on a per floorspace basis and supports regional assessments of energy efficiency, decarbonization potential, and the cost effectiveness of building renovations over time.

There is an urgent need for a systematic, high-resolution, and openly accessible global building floorspace database, particularly for data-scarce emerging economies. This effort should integrate satellite imagery, statistical modeling, and bottom-up physical surveys to improve spatial and temporal granularity. Future research should prioritize multidimensional building data to support advanced stock assessments and facilitate the integration of smart and clean technologies on rooftops and facades, such as BIPVs and distributed energy systems.

Moreover, a paradigm shift is needed—from purely technological advancements to sufficiency-based strategies. Greater attention should be paid to building sufficiency, such as developing tiny houses with downsizing and simple living, improving the renovation rate of old buildings to extend their lifetime and avoid unnecessary reconstruction, and considering co-housing options within neighborhoods. When paired with the LCA and globally consistent data, sufficiency-oriented design can provide a stronger foundation for the sustainable and equitable decarbonization of the built environment.

## EXPERIMENTAL PROCEDURES

### Resources availability

#### *Lead contact*

Further information and requests for resources should be directed to and will be fulfilled by the lead contact, Dr. Nan Zhou (nzhou@lbl.gov).

#### *Data and code availability*

The data calculated in this study has been deposited in the GitHub repository, as well as https://doi.org/10.1016/j.ynexs.2024.100019. This study did not generate new unique code.

#### *Materials availability*

This study did not generate new unique materials.

### Overview of the bibliometric analysis

This review focused on building floorspace and the building stock, as they are key factors influencing carbon emission intensity across the full life cycle of buildings and determining the total carbon emissions of the building sector. For the literature search, we used the WOS search engine to explore relevant studies. Table S1 in the Supplemental Information outlines the query set applied in this process. The “OR” logic operator was used to connect synonymous keywords within each query set, whereas the “AND” logic operator was used to link different query sets. Additionally, Figure S1 in the Supplemental Information illustrates the specific steps followed during the search.

## SUPPLEMENTAL INFORMATION

The supplemental materials are included at the end of this submission file.

- Document S1: A PDF file that includes Figure S1 and Table S1.
- Table S2: An Excel file containing additional data that is too large to fit in the PDF, related to Table 1.

## ACKNOWLEDGMENTS

First author appreciates the National Planning Office of Philosophy and Social Science Foundation (24BJY129). Corresponding author declares that this manuscript was authored by an author at Lawrence Berkeley National Laboratory under Contract No. DE-AC02-05CH11231 with the U.S. Department of Energy. The U.S. Government retains, and the publisher, by accepting the article for publication, acknowledges, that the U.S. Government retains a non-exclusive, paid-up, irrevocable, world-wide license to publish or reproduce the published form of this manuscript, or allows others to do so, for U.S. Government purposes.

## AUTHOR CONTRIBUTIONS

Conceptualization, M.M., N.Z., and J.Y.; Methodology, M.M., S.Z., J.L., and R.Y.; Software, S.Z., and J.L.; Validation, M.M., R.Y., and W.C.; Writing-Original Draft, M.M., and S.Z.; Writing-Review & Editing, M.M., N.Z., J.Y., R.Y., and W.C.; Funding Acquisition, M.M., and N.Z.; Supervision, M.M., N.Z., and J.Y..

## DECLARRATION OF INTERESTS

## FIGURE TITLES AND LEGENDS

**Figure 1.** Global trends in building floorspace. (A) Global and regional building stock growth rates, 2020-2070; (B) regional stock distribution in 2070. Note: EU27 refers to the 27 countries that are part of the European Union. The classification of developed and developing economies is based on the Human Development Index defined by the United

Nations Development Programme (source: *https://hdr.undp.org/data-center/human-development-index#/indicies/HDI*).

**Figure 2.** Meta-analysis of the retrieved articles. (A) Distribution of publication countries/regions; (B) institution cooperation network; (C) journals with more than ten publications; and (D) the keyword cloud.

**Figure 3.** The hierarchical structure of building floorspace applications in the retrieved articles, categorized by building types, life cycle stages, research topics and methods, and goals. Note: the width of each area reflects the number of articles associated with each category.

**Figure 4.** Schematic diagram of the building stock turnover analysis process.

**Figure 5.** Trends in per capita floorspace for (A) residential and (B) non-residential buildings worldwide, 2000-2022.

**Figure 6.** Growth trends in per capita floorspace for (A) residential and (B) non-residential buildings worldwide under the BAU scenario, 2022-2070; (C) the global building stock level in 2070.

## FORMULAE AND EQUATIONS

None.

## TABLE TITLES AND LEGENDS

**Table 1.** Summary of the top 10 most-cited review articles from our retrieved literature on building floorspace and the building stock in relation to energy use and carbon emissions. Note: Citation data are based on the Essential Science Indicators (ESI) and the WOS Core Collection, accessed on 03/15/2025.

## STANDARDIZED DATA REPORTING

None.

## SUPPLEMENTAL INFORMATION DESCRIPTION

The supplemental information includes one PDF file and one Excel file.